\documentclass[aps, prl, showpacs, reprint, superscriptaddress]{revtex4-1}

\usepackage{color}
\usepackage{graphicx}
\usepackage{dcolumn}
\usepackage{bm}

\begin{document}

\title{Concentration of Vacancies at Metal Oxide Surfaces: Case Study of MgO (100)}

\author{Norina A. Richter}
\affiliation{Fritz-Haber-Institut der Max-Planck-Gesellschaft, Faradayweg 4-6, 14195 Berlin, Germany}
\author{Sabrina Sicolo}
\affiliation{Institut f\"ur Chemie, Humboldt Universit\"at zu Berlin, Unter den Linden 6, 10099 Berlin, Germany}
\author{Sergey V. Levchenko}
\affiliation{Fritz-Haber-Institut der Max-Planck-Gesellschaft, Faradayweg 4-6, 14195 Berlin, Germany}
\author{Joachim Sauer}
\affiliation{Institut f\"ur Chemie, Humboldt Universit\"at zu Berlin, Unter den Linden 6, 10099 Berlin, Germany}
\author{Matthias Scheffler}
\affiliation{Fritz-Haber-Institut der Max-Planck-Gesellschaft, Faradayweg 4-6, 14195 Berlin, Germany}

\begin{abstract}
We investigate effects of doping on formation energy and concentration of oxygen vacancies at a metal oxide surface, using MgO (100) as an example. Our approach employs density-functional theory, where the performance of the exchange-correlation functional is carefully analyzed, and the functional is chosen according to a fundamental condition on DFT ionization energies. The approach is further validated by CCSD(T) calculations for embedded clusters. We demonstrate that the concentration of oxygen vacancies at a doped oxide surface is largely determined by formation of a macroscopically extended space charge region. 
\end{abstract}

\pacs{61.72.Bb, 61.72.jd, 68.55.Ln, 68.47.Gh, 68.35.-p}

\keywords{vacancies, defect concentration, defect formation energy,
doping concentration, space charge, band bending, metal oxide surface,
defect-defect interaction, density-functional-theory,
ab initio atomistic thermodynamics, magnesium oxide}
\maketitle

\begin{figure}[!hbt]
\includegraphics[width=1\columnwidth]{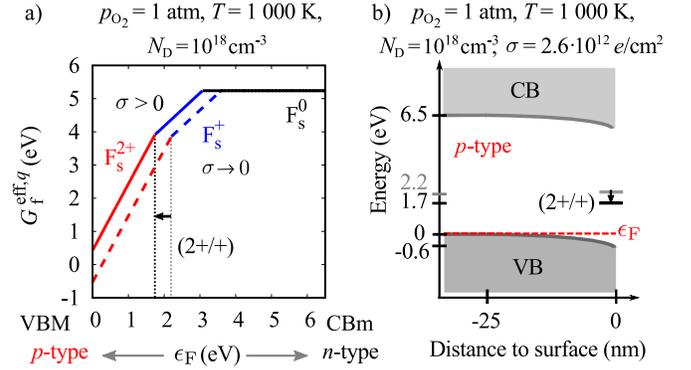}
\caption{a) Calculated Gibbs free energies of formation $G_{\rm f}^{\rm eff,\it q}$ (see text) of F centers at MgO (100) for $T=1~000$~K and normal pressure of oxygen, as a function of Fermi energy, $\epsilon_{\rm F}$, between valence-band maximum, VBM, and conduction band minimum, CBm. Realistic dopant concentration $N_{\rm D}$=10$^{18}$~cm$^{-3}$ and surface charge density $\sigma=2.6\cdot10^{12}\frac{e}{{\rm cm}^2}$ (solid lines) and the dilute limit $\sigma\rightarrow0$ (dashed lines) are shown.
b) In $p$-type MgO ($\epsilon_{\rm F}$=VBM) under realistic conditions, band bending, due to formation of a space charge region, limits the formation of surface F$_{\rm s}^{2+}$ centers.} 
\label{fig:bb} 
\end{figure}

Metal oxides are key materials for many technological applications. For example, MgO is used as a catalyst for methane oxidation, TiO$_2$ plays an important role as a photocatalyst, and RuO$_2$ catalyzes the oxidation of carbon monoxide. 
It is generally accepted that intrinsic point defects, in particular oxygen vacancies (also called F or color centers), play a decisive role in catalysis at oxide surfaces~\cite{{balint_specific_2001},{wahlstrom_electron_2004},{yan_role_2005},{scanlon_competing_2009}},
but significant controversy exists regarding their formation energy, concentration, and charge state. In this paper we study these issues for MgO bulk and the MgO (100) surface in contact with an O$_2$ gas phase at realistic temperature and pressure. Furthermore, we consider that realistic metal oxides are typically doped, either intentionally or unintentionally. Although the experimental band gap of MgO is 7.8~eV~\cite{whited_exciton_1969}, 
realistic samples are typically neither clear transparent nor insulating. Defects such as intrinsic point defects, impurities, and defect complexes can give rise to electron or hole conductivity~\cite{mitoff_electrical_1959, balint_specific_2001, hadi_electrical_2012}. 
In this paper, we neglect defect complexes (e.g. dopant plus vacancy). Thus, for our study the role of dopants is to create a Fermi level, {\em i.e.} a reservoir for electrons and holes. This is termed ``the global effect of doping``. We focus our discussion on $p$-type MgO, since it exhibits intriguing catalytic properties~\cite{balint_specific_2001,dubois_common_1990, arndt_critical_2011}. Still our theoretical model covers also $n$-type material, where the concentration of F centers is very low. 
Our main finding is that for $p$-type samples, surface O vacancies get doubly positively charged which lowers their formation energy and results in a significant defect concentration.
In fact, the free energy of formation in thermodynamic equilibrium is negative under typical catalytic temperatures and pressures. We show that the limiting factor to formation of surface oxygen vacancies is the formation of a space charge region (Fig.~\ref{fig:bb}). Although well-known for doped semiconductor surfaces, the effect of space charge and band bending on the concentration of surface defects has not been discussed so far.   

The values of calculated defect energy levels and total energies are sensitive to the employed treatment of exchange and correlation (xc) of the many-electron system. Therefore, special attention is given below to this aspect: Our approach is to determine the best xc functional of the HSE family by the condition that the ionization potentials obtained with the optimized (opt-) HSE functional agree with the results of a $G_0 W_0$ calculation. In exact DFT such condition must be fulfilled exactly~\cite{levy_approach_1996}. However, due to the limited flexibility of the HSE family of functionals, which range from the PBE~\cite{perdew_generalized_1996} generalized gradient approximation via the HSE06~\cite{krukau_influence_2006} hybrid functional to the PBE0~\cite{perdew_rationale_1996} hybrid functional, our approach is obviously not exact, but it is ``the best compromise``. In terms of the HSE~\cite{heyd_hybrid_2003} exchange and range-separation parameters ($\alpha, \omega$), the three mentioned functionals correspond to (0, arbitrary), (0.25, 0.11 bohr$^{-1}$), (0.25, 0). Our approach is validated for neutral, embedded Mg$_x$O$_y$ clusters using the CCSD(T) method.

In this work, we use the all-electron FHI-aims code~\cite{blum_ab_2009-1} for the periodic structures (bulk and surfaces) and some embedded cluster calculations.
FHI-aims employs atom-centered numeric basis functions and various xc functionals as well as the $GW$ approach.
The basis set and numerical grids are of high quality as defined by the \textit{tight}~\cite{blum_ab_2009-1} settings. 
For all periodic surface models full atomic relaxation is calculated using PBE at the PBE bulk lattice parameter (4.258~\AA). HSE calculations are performed at these geometries, since using HSE geometries for two smallest unit cells considered results in negligible changes in the calculated formation energies.
Vibrational energy $\Delta F^q_{\rm vib}(T)$ and vdW contributions to the formation energies were analyzed as well~\cite{to_be_publ}, but found to be insignificant for this study.
 
Furthermore, we employ the TURBOMOLE program package~\cite{turbomole} for various embedded clusters using different xc functionals and CCSD(T). 
Triple-zeta valence plus polarization basis sets [$5s3p2d1f$] / [$5s4p3d$] are used~\cite{weigend_2005}. For the CCSD(T) computations we correlate also electrons in the Mg $2s$ and $2p$ shells, using core-valence correlation consistent basis sets, cc-pCVXZ (X = D, T, Q). On the O$^{2-}$ ions we use the aug-cc-pVXZ basis sets~\cite{dunning}. In both CCSD(T) and DFT calculations, the basis set superposition error was evaluated following the Boys-Bernardi counterpoise correction~\cite{boys_1970}. MgO clusters are embedded in a periodic point charge array using the periodic electrostatic embedded cluster model~\cite{burow_2009} in TURBOMOLE and a converged finite set of point charges in FHI-aims. (See supplemental information (SI) at [URL will be inserted by publisher] for more details.) 

To minimize non-physical polarization of peripheral oxygen anions by the embedding point charges, pseudopotentials were added to the first shell of embedding Mg$^{2+}$ cations (all-electron Hay\&Wadt effective core potentials~\cite{hay_1985} in TURBOMOLE, and Troullier-Martins-type norm-conserving non-local pseudopotentials~\cite{kleinman_efficacious_1982, fuchs_ab_1999} in FHI-aims). The PBE lattice constant has been used for the embedded clusters.
Apart from the outermost frozen shell of atoms, full relaxation is allowed for in the cluster calculations, except for the CCSD(T) and $GW$ calculations and respective DFT values.

In realistic samples defects may get charged due to electron transfer between the dopant-induced Fermi level and the defect states~\cite{weinert_defects_1986}. The neutral oxygen vacancy in MgO bulk and at the MgO (100) surface has an energy level deep in the band gap. This state has $s$-like symmetry at the defect site, and is fully occupied by two electrons. Thus, a singly and even a doubly charged vacancy is possible. In FHI-aims this situation can be modeled in two ways: either by adding a constant charge density to the density entering the Hartree term, or by slightly modifying the nuclear charges of the atoms in the unit cell. Either concept enables us to describe a charged defect while the supercell is kept neutral. The ``constant density approach`` is the standard treatment in other periodic codes. For surfaces, where much of the supercell corresponds to vacuum, this approach is obviously unphysical, although it can be partially remedied by {\em a posteriori} correction schemes~\cite{komsa_finite-size_2013}. The other treatment corresponds to the virtual-crystal approximation (VCA)~\cite{vegard_1921, scheffler_lattice_1987} of a crystal with dopants. 
We change the nuclear number of all Mg atoms in the supercell by $\Delta Z_{\rm Mg} = -q/N_{\rm Mg}$,
where $q$ is the charge of the oxygen vacancy (+1 or +2), and $N_{\rm Mg}$ is the number of Mg atoms in the supercell. This means that 1 or 2 electrons are transferred to the VBM, which is the Fermi level in the virtual crystal. 
Once known for one particular Fermi level, the O vacancy formation energy can be trivially calculated for an arbitrary Fermi level (see Eq.~\ref{eq:GIBBS_FORM}).    

When removing atoms from the bulk or from the surface of a material, we need to consider also a reservoir to which the atoms are brought~\cite{scheffler_parameter-free_1988, reuter_ab_2005}. We assume a gas phase of O$_2$ molecules which is characterized by an oxygen chemical potential, $\mu_{\rm O}(T,p)$~\cite{reuter_ab_2005}. For an isolated oxygen vacancy, the Gibbs free energy of formation is:
\begin{equation}
G_{\rm f}^{q}=E_{\rm vac}^{q}-E_{\rm host}+\mu_{\rm O}+q\epsilon_{\rm F}+\Delta F^q_{\rm vib}(T).
\label{eq:GIBBS_FORM}
\end{equation}
Here, $E_{\rm vac}^{q}$ and $E_{\rm host}$ are DFT total energies of defected and undefected systems, respectively, $\Delta F^q_{\rm vib}$ is the change in vibrational Helmholtz free energy of the crystal upon defect formation, $q$ is the defect charge, and $\epsilon_{\rm F}$ is the Fermi energy. The oxygen chemical potential is
\begin{equation}
\mu_{\rm O} = E_{\rm O} - \frac{1}{2} E_{\rm O_2}^{\rm bind} + \Delta \mu_{\rm O},
\label{eq:mu-O}
\end{equation}
where $\Delta \mu_{\rm O}$ contains the vibrational and other $T$- and $p$-dependent terms~\cite{reuter_ab_2005}. We use the experimental binding energy without zero point energies
$E_{\rm O_2}^{\rm bind}=5.22$~eV~\cite{feller_re-examination_1999} to reduce artifacts originating from the generalized gradient approximations for the binding of the O$_2$ molecule, but the calculated total energy of the free atom, $E_{\rm O}$, is calculated with the corresponding electronic-structure approach. $\Delta \mu_{\rm O} = 0$ defines the oxygen-rich limit.

First, we address formation energies for isolated vacancies in the bulk, $G_{\rm f}^{\rm bulk,\it q}$. We extrapolate our DFT formation energies to the dilute limit using Taylor expansion in terms of reciprocal supercell lattice constant. In agreement with related work~\cite{ramprasad_new_2012}, we find that the charge-transition levels (2+/+) and (+/0), as well as formation energies $G_{\rm f}^{\rm bulk,\it q}$, are almost independent on the xc functional within the HSE family, when referenced to the vacuum level. 
However, given the more realistic situation of $p$-type material, where $\epsilon_{\mathrm{F}}$ is at VBM, $G_{\rm f}^{\rm bulk,\it q}$ does depend strongly on the choice of HSE parameters ($\alpha$, $\omega$) for $q\neq 0$. The formation energy of the neutral defect is insensitive to the functional. We find that the main error in charged defect formation energies is the error in the VBM position with respect to vacuum (also pointed out in~\cite{west_importance_2012}). For fixed $\omega$ the formation energies depend practically linearly on the exchange parameter $\alpha$, which can be traced back to a linear dependence of VBM with respect to vacuum on $\alpha$. 

Next, we identify the optimal xc functional to describe the formation energies of F centers in MgO. 
The ionization potential at a fixed defect geometry for a given functional HSE($\alpha$, $\omega$) is 
\begin{equation}
I^{q\rightarrow q+1}_{\Delta {\rm SCF}}=E_{\rm vac}^{q+1}+\epsilon_{\rm F}-E_{\rm vac}^{q},
\label{eq:SCF_IP}  
\end{equation}
where both $E_{\rm vac}^{q}$ and $E_{\rm vac}^{q+1}$ are extrapolated to the dilute limit. For $\epsilon_{\rm F}=$VBM,
$I^{q\rightarrow q+1}_{\Delta {\rm SCF}}$ depends on ($\alpha$, $\omega$).
In exact DFT the Kohn-Sham highest occupied orbital (HOMO) does not change with occupation and agrees with the ionization energy. A more practical request is that the HOMO, calculated by $G_0W_0$ on top of the HSE electronic ground state, should agree with the HSE ionization energy
\begin{equation}
I^{q\rightarrow q+1}_{G_0W_0}=\epsilon_{\rm F}-\epsilon^{G_0W_0}_{\rm HOMO}\stackrel{!}{=}I^{q\rightarrow q+1}_{\Delta {\rm SCF, opt-\rm HSE}},
\label{eq:CROSS_AFFINITY}
\end{equation}
identifying what we call the optimized HSE functional, opt-HSE that correctly describes the charge excitation of the defect.

\begin{figure}[!htb]
\begin{center}
\includegraphics[width=0.7\columnwidth]{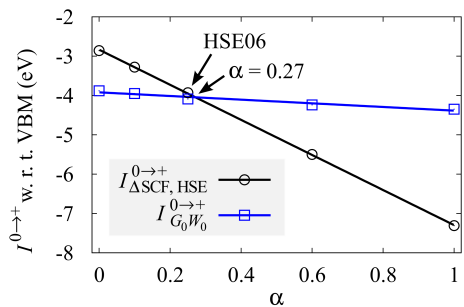}
\caption{Ionization potential at F$^0$ geometry calculated for an Mg$_6$O$_9$ embedded cluster by $\Delta$SCF with HSE xc functionals (black symbols) and from the HOMO of a $G_0W_0$@HSE calculation (blue symbols). The screening parameter is $\omega=0.11$ bohr$^{-1}$. Solid lines show linear fits to the ionization energies as a function of exchange parameter $\alpha$.}
\label{fig:GW}
\end{center} 
\end{figure}

We determine the opt-HSE functional for fixed $\omega=0.11$ bohr$^{-1}$. The ionization energy $I^{0\rightarrow +}$ for $\epsilon_{\rm F}$ at VBM at F$^0$ geometry is calculated for an embedded Mg$_6$O$_9$ cluster model using FHI-aims. The Fermi level $\epsilon_{\rm F}$ is obtained as ${\rm VBM}=E_{\rm host}^{+1}-E_{\rm host}$ using HSE, and from the HOMO of the host system in the corresponding $G_0W_0@{\rm HSE}$ calculations. The ionization potential shows a near-linear dependence on the exchange parameter $\alpha$ (Fig.~\ref{fig:GW}) for both $\Delta$SCF and $GW$ method. The intersection of the two linear fits is at $\alpha$=0.27, very close to HSE06 with parameter set ($\alpha$=0.25, $\omega=0.11$ bohr$^{-1}$). We therefore use HSE06 as our opt-HSE functional that correctly describes the charge excitation of the defect. The difference in formation energies with $\alpha$=0.25 instead of $\alpha$=0.27 is negligible for F$^0$, less than 0.1~eV for F$^+$, and less than 0.2~eV for F$^{2+}$.

We perform a validation for the F$^0$ formation energy using an unrelaxed Mg$_6$O$_9$ cluster model and the CCSD(T) method. This results in a correction $\Delta$CCSD(T) of the DFT formation energies of $-0.09$~eV for PBE, $0.07$~eV for PBE0, and $-0.28$~eV for B3LYP. Adding these corrections to the DFT formation energies ($\Delta{\mu_{\rm O}}=0$) for a converged cluster size Mg$_{16}$O$_{19}$ yields DFT+$\Delta$CCSD(T) results of 6.85, 6.88 and 6.89 eV, respectively. These numbers are in good agreement with the HSE06 F$^0$ formation energies 7.04 eV and 7.05 eV obtained from the same converged embedded cluster and periodic calculations, respectively, using FHI-aims.

Thus, HSE06 is the opt-HSE functional in accordance with $GW$ as well as coupled-cluster results. 
Our results show that the experimental value for the bulk F$^0$ center formation enthalpy in MgO of 9.29~eV with respect to the O$_2$ molecule~\cite{kappers_f^+_1970, weast_crc_1980} is a significant overestimate. A likely reason is that thermodynamic equilibrium has not been reached in this experiment. 

We are now on solid grounds to provide an accurate estimate of $G^{{\rm surf,} q}_{\rm f}$ for isolated oxygen vacancies at the surface using our periodic model and the HSE06 xc functional. F$^0_{\rm s}$, F$^+_{\rm s}$, and F$^{2+}_{\rm s}$ formation energies in the dilute limit are 6.34~eV, 2.76~eV, and 0.55~eV, respectively, for $\Delta{\mu_{\rm O}}=0$ and $\epsilon_{\rm F}$ at VBM. For more realistic conditions, the formation energies are lower, as shown in Fig.~\ref{fig:bb}a).
$E^{\rm surf,0}_{\rm f}$ obtained with the CCSD(T) correction method, is, as for the bulk, in good agreement with the HSE06 value. The corrections $\Delta$CCSD(T) to the DFT formation energies, calculated with an unrelaxed Mg$_5$O$_5$ model, are $-0.26$~eV for PBE, $-0.01$~eV for PBE0, and $-0.28$~eV for B3LYP, yielding DFT+$\Delta$CCSD(T) values of 6.23, 6.25 and 6.33~eV, respectively.

Formation energies of {\em neutral} O vacancies depend weakly on their concentration (up to approx. 3\% for bulk and 12\% for surface defects in MgO). On the contrary, due to the slow decay of Coulomb interaction with distance, the formation energy of {\em charged} defects will strongly depend on their concentration, as well as the distribution of the compensating charge. Thus, concentration of dopants $N_{\rm D}$ and their distribution (doping profile) should have a strong global effect on the defect formation energies. The equilibrium concentrations $\eta_q$ of oxygen vacancies in three different charge states ($q=0-2$) are determined by the minimum of the total free energy $G$ of the system with interacting defects:
\begin{equation}
\frac{\partial G}{\partial \eta_q}=G_{\rm f}^{{\rm eff,}q}(\eta_0,\eta_1,\eta_2)-T\frac{\partial s_{\rm conf}(\eta_0,\eta_1,\eta_2)}{\partial \eta_q}=0,
\label{eq:GTOT}
\end{equation}
where
\begin{equation}
G_{\rm f}^{{\rm eff,}q}(\eta_0,\eta_1,\eta_2)= \frac{\partial}{\partial\eta_q}\sum_{r=0}^2\eta_{r} G_{\rm f}^{r}(\eta_0,\eta_1,\eta_2)
\label{eq:GEFF}
\end{equation}
is an effective formation energy of a vacancy in charge state $q$ in the presence of other vacancies. The configurational entropy $s_{\rm conf}$ accounts for energetically degenerate arrangements of the defects (see SI).

Surface defects are charged by accommodating charge carriers from the bulk. This results in depletion of the charge carriers and creation of a space charge layer in the subsurface region. The resulting electrostatic potential causes band bending and prevents more charges from the bulk to reach the surface, increasing the energy cost per defect. As a result, there are two leading electrostatic contributions to the formation energy of charged defects: (i) attraction to the compensating charge, and (ii) band bending. The first contribution reduces the formation energy, while the second contribution increases it. The thickness of the space charge layer, $z^{\rm SC}$, depends on the doping profile, and may be limited by the thickness of the material. We consider the case of uniformly distributed dopants and unconstrained $z^{\rm SC}$, but the discussion can be straightforwardly generalized to the more complex situations. To stay focussed on electrostatic effects, we also do not consider a possible ($T$,$p$) dependence of the bulk charge carrier density $N^{\rm bulk}_{e/h}$, {\em i.e.} $N^{\rm bulk}_{e/h}\equiv N_{\rm D}$ is a constant external parameter.

The dependence of $G_{\rm f}^{{\rm surf,}q}$ on the surface charge density $\sigma$ is calculated as follows. First, we calculate formation energies $G_{\rm f}^{{\rm surf,}q}(\sigma ,z^{\rm SC})$ at a fixed $z^{\rm SC}$, equal to the slab thickness $d$, using VCA (see SI). The calculated formation energies include both electrostatic effects mentioned above. The actual $z^{\rm SC}$ is determined by $N_{\rm D}$ as follows:
\begin{equation}
z^{\rm SC} = \frac{\sigma }{e N_{\rm D}},
\label{eq:Z}
\end{equation}
where $e$ is the absolute value of the electron charge.
The formation energy as a function of $\sigma$ and $z^{\rm SC}$ is
\begin{equation}
G_{\rm f}^{{\rm surf,}q}(\sigma ) = G_{\rm f}^{{\rm surf,}q}(\sigma ,d)-qE^{\rm SC}(\sigma ,d)+qE^{\rm SC}(\sigma ,z^{\rm SC}),
\label{eq:ESIG}
\end{equation}
where 
\begin{equation}
E^{\rm SC}(\sigma ,z^{\rm SC})=\frac{e\sigma}{6\epsilon_{\rm r}\epsilon_0}z^{\rm SC}
\end{equation}
is the classic expression for the energy of the space charge region formation at a semiconductor surface. The temperature dependence of $z^{\rm SC}$ and $E^{\rm SC}(\sigma ,z^{\rm SC})$ at fixed $\sigma$ is neglected. The meaning of the last two terms in Eq.~\ref{eq:ESIG} is to replace the band bending contribution to the formation energy calculated for $z^{\rm SC}=d$ with the one obtained for the actual $z^{\rm SC}$ from Eq.~\ref{eq:Z}. The remaining dependence on $\sigma$ after subtracting $qE^{\rm SC}(\sigma ,d)$ from $G_{\rm f}^{{\rm surf,}q}(\sigma ,d)$ is due to the electrostatic attraction between the defect and the compensating charge. 

\begin{figure}[!bht]
\begin{center}
\includegraphics[width=1\columnwidth]{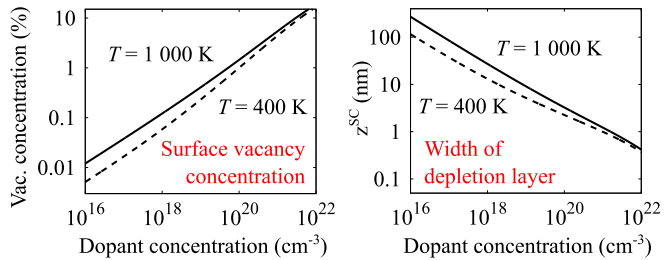}
\caption{{\em Left}: F$_{\rm s}^{2+}$ concentration as a function of dopant concentration $N_{\rm D}$ for two different temperatures and oxygen partial pressure of 1~atm. {\em Right}: Dependence of the space charge depth $z^{\rm SC}$ on $N_{\rm D}$.}
\label{fig:conc}
\end{center} 
\end{figure}

We can now calculate the equilibrium concentration of O vacancies at a $p$-doped MgO (100) surface, using Eq.~\ref{eq:GTOT}. $G_{\rm f}^{{\rm eff,}q}(\sigma)$ is calculated from Eq.~\ref{eq:GEFF} with $\sigma = e\eta_1+2e\eta_2$. The concentrations of F$^0_{\rm s}$ and F$^+_{\rm s}$ are found to be negligible at realistic $T$, $p_{\rm O_2}$, and $N_{\rm D}$. The F$_{\rm s}^{2+}$ concentration $\eta_2$ and corresponding $z^{\rm SC}$ as functions of $N_{\rm D}$ are shown in Fig.~\ref{fig:conc} for different temperatures and $p_{\rm O_2}=1$~atm. Although the F$^{2+}_{\rm s}$ Gibbs free energy of formation at $\sigma\rightarrow 0$ is small or even negative at elevated temperatures, the equilibrium defect concentration does not exceed $\sim 0.5\%$ at $N_{\rm D}\leq 10^{18}$~cm$^{-3}$. Thus, space charge layer formation can be a mechanism by which wide-band-gap semiconductor surfaces remain stable at high temperatures.
The band bending profile for $\epsilon_{\rm F}={\rm VBM}$ at $T=1~000$~K for $p_{\rm O_2}=1$~atm and $N_{\rm D}$=10$^{18}$~cm$^{-3}$ is shown in Fig.~\ref{fig:bb}b). Under these conditions the bulk bands bend downwards by 0.6~eV, and the (2+/+) charge transition level is lowered from 2.2~eV to 1.7~eV above the Fermi level. At $T=1~000$~K and $p_{\rm O_2}=1$~atm, the contribution of the electrostatic attraction term is small for small $N_{\rm D}$, but becomes comparable to the formation energy in the dilute limit for $N_{\rm D}>10^{18}$~cm$^{-3}$.

We have presented a methodology for calculating charged defect formation energies and concentrations at surfaces, taking into account electrostatic effects due to charge transfer between surface and bulk. Doped material has been simulated using the VCA, and an optimal DFT functional has been identified by validation with coupled-cluster and \textit{GW} methods. Our analysis shows that the concentration of F$_{\rm s}^{2+}$ centers at the (100) terrace of $p$-type MgO can be as high as 1\% at realistic conditions, while relative F$_{\rm s}^{+}$ and F$_{\rm s}^0$ concentrations are negligible. We demonstrate that the formation of charged vacancies creates a localized, although macroscopically extended, space charge region. This decreases charge transition levels with respect to Fermi level at the surface, raising the formation energy by up to 1~eV and, therefore, limiting the defect concentration. 
 We conclude that electrostatic effects can largely control oxygen vacancy formation at the surface of metal oxides. Experimental information on doping profiles may provide new insights on catalytic activity of doped oxide surfaces.

\begin{acknowledgments}
This collaboration was financially supported by the cluster of excellence UniCat. N. A. R. acknowledges financial support from the International Max Planck Research School "Complex Surfaces in Materials Science". S. V. L. acknowledges financial support from the Alexander von Humboldt-Foundation.
\end{acknowledgments}

\end{document}